\documentclass{elsart}

\usepackage[square,comma]{natbib}
\usepackage{graphicx}
\usepackage{pxfonts}
\usepackage{lineno}

\usepackage{amssymb}
\usepackage{footnote}
\usepackage{multirow}
\usepackage{varwidth}
\usepackage{array}

\journal{}

\begin{document}

\thispagestyle{empty}
\begin{Large}
\textbf{DEUTSCHES ELEKTRONEN-SYNCHROTRON}

\textbf{\large{Ein Forschungszentrum der Helmholtz-Gemeinschaft}\\}
\end{Large}

DESY 13-109

June 2013

\begin{eqnarray}
\nonumber
\end{eqnarray}
\begin{center}
\begin{Large}
\textbf{Extension of SASE bandwidth up to $2 \%$ as a way to
increase the efficiency of protein structure determination by x-ray
nanocrystallography \\ at the European XFEL}
\end{Large}
\begin{eqnarray}
\nonumber &&\cr \nonumber
\end{eqnarray}

\begin{large}
Svitozar Serkez$^a$, Vitali Kocharyan$^a$, Evgeni Saldin$^a$, Igor
Zagorodnov$^a$, Gianluca Geloni$^b$, and Oleksander Yefanov$^c$
\end{large}

\textsl{\\$^a$Deutsches Elektronen-Synchrotron DESY, Hamburg}
\begin{large}

\end{large}
\textsl{\\$^b$European XFEL GmbH, Hamburg}
\begin{large}

\end{large}
\textsl{\\$^c$Center for Free-Electron Laser Science, Hamburg}
\begin{eqnarray}
\nonumber
\end{eqnarray}
\begin{eqnarray}
\nonumber
\end{eqnarray}
ISSN 0418-9833
\begin{eqnarray}
\nonumber
\end{eqnarray}
\begin{large}
\textbf{NOTKESTRASSE 85 - 22607 HAMBURG}
\end{large}
\end{center}
\clearpage
\newpage

\begin{frontmatter}



\title{Extension of SASE bandwidth up to $2 \%$ as a way to increase the efficiency of protein structure determination by x-ray nanocrystallography \\ at the European XFEL}


\author[DESY]{Svitozar Serkez \thanksref{corr},}
\thanks[corr]{Corresponding Author. E-mail address: svitozar.serkez@desy.de}
\author[DESY]{Vitali Kocharyan,}
\author[DESY]{Evgeni Saldin,}
\author[DESY]{Igor Zagorodnov,}
\author[XFEL]{Gianluca Geloni,}
\author[CFEL]{and Oleksandr Yefanov}

\address[DESY]{Deutsches Elektronen-Synchrotron (DESY), Hamburg, Germany}
\address[XFEL]{European XFEL GmbH, Hamburg, Germany}
\address[CFEL]{Center for Free-Electron Laser Science, Hamburg, Germany}

\begin{abstract}
Femtosecond x-ray nanocrystallography  exploiting XFEL radiation is
an emerging method for protein structure determination using
crystals with sizes ranging from a few tens to a few hundreds
nanometers. Crystals are randomly hit by XFEL pulses, producing
diffraction patterns at unknown orientations. One can determine
these orientations by studying the diffraction patterns themselves,
i.e. by indexing the Bragg peaks. The number of indexed individual
images and the SASE bandwidth are inherently linked, because
increasing the number of Bragg peaks per individual image requires
increasing the bandwidth of the spectrum. This calls for a few
percent SASE bandwidth, resulting in an increase in the number of
indexed images at the same number of hits. Based on start-to-end
simulations for the baseline of the European XFEL, we demonstrate
here that it is possible to achieve up to a tenfold increase in SASE
bandwidth, compared with the nominal mode of operation. This
provides a route for further increasing the efficiency of protein
structure determination at the European XFEL. We illustrate this
concept with simulations of lysozyme nanocrystals.
\end{abstract}

%
%
%
\end{frontmatter}



\section{\label{sec:intro} Introduction}

X-ray crystallography is currently the leading method for imaging
macromolecules with atomic resolution. Third generation synchrotron
sources allow for successful structure determination of proteins.
The size of a typical single crystal used for conventional protein
crystallography is in the order of $50 \mu$m $- 500 \mu$m
\cite{HUNT}. Obtaining sufficiently large crystals is currently a
serious stumbling block as regards structure determination. The new
technique of femtosecond nanocrystallography is based on data
collection from a stream of nanocrystals, and ideally fills the gap
between conventional crystallography, which relies on the use of
large, single crystals, and single-molecular x-ray diffraction.

The availability of XFELs allows for a new "diffraction before
destruction" approach to overcome radiation damage due to the
ultrafast and ultrabright nature of the x-ray pulses, compared to
the time scale of the damage process \cite{HAJD}-\cite{SEIB}. In
fact, if such time-scale is longer than the pulse duration, the
diffraction pattern yields information about the undamaged material.
Femtosecond nanocrystallography involves sequential illumination of
many small crystals of proteins by use of an XFEL source
\cite{CHA3}. The high number of photons incident on a specimen are
expected to produce measurable diffraction patterns from
nanocrystals, enabling structure determination with high resolution
also for systems that can only be crystallized into very small
crystals and are not suitable, therefore, for conventional
crystallography. Each crystal is used for one exposure only, and the
final integrated Bragg intensities must be constructed from
"snapshot" diffraction patterns containing partially recorded
intensities. Each pattern corresponds to a different crystal at
random orientation. Experiments at the LCLS \cite{LCLS2} confirmed
the feasibility of the "diffraction before destruction" method at
near atomic resolution using crystals ranging from $0.2
 ~\mu$m to $3 ~\mu$m \cite{CHA3}. This method relies on x-ray SASE pulses with a
few mJ energy, a few microradians angular spread, and about $0.2 \%$
bandwidth with a photon energy range between 2 keV and 9 keV.

The success of nanocrystallography depends on the robustness of the
procedure for pattern determination. After acquisition, diffraction
patterns are analyzed to assign indexes to Bragg peaks (indexing
procedure). Each of the indexed peaks is integrated in order to
obtain an intensity. Intensities of corresponding peaks are averaged
within the dataset. The table of peak indexes and intensities
obtained in this way is used for protein structure determination.
Indexing algorithms used in crystallography enable to determine the
orientation of the diffraction data from a single crystal when a
relatively large number of reflections are recorded. Femtosecond
nanocrystallography brings new challenges to data processing
\cite{KIRI}. The problem is that single snapshots of crystal
diffraction patterns may contain very few reflections,  which are
not enough for indexing. In this paper we will show how to overcome
this obstacle.

The number of Bragg peaks is proportional to the bandwidth of the
incident radiation pulse. Considering the baseline configuration of
the European XFEL \cite{TSCH}, and based on start-to-end
simulations, we demonstrate here that it is possible to achieve a
tenfold increase in bandwidth by strongly compressing electron
bunches with a charge of $0.25$ nC up to 45 kA. This allows data
collection with a $2 \%$ bandwidth, a few mJ radiation pulse energy,
a few fs pulse duration, and a photon energy range between 2 keV and
6 keV, which is the most preferable range for nanocrystallography
\cite{BERG}.

The generation of x-ray SASE pulses at the European XFEL using
strongly compressed electron bunches has many advantages, primarily
because of the very high peak power and very short pulse duration
that can be achieved in this way \cite{DOHL}. Here we demonstrate
(see Section \ref{sec:FELstudy} for more details) that a few TW
power-level can be achieved in the SASE regime, and with 3 fs-long
pulses. For electron bunches with very high peak-current, the
wakefields (mainly due to the undulator vacuum pipe) have a very
important effect on the lasing process. In fact, the variation in
energy within the electron bunch due to wakefield effects is large,
and yields, as a consequence, a tenfold increase in the SASE
radiation bandwidth. However, as we demonstrate here (see Section
\ref{sec:pot} in more details), a large bandwidth presents many
advantages for nanocrystallography, and provides a route for
increasing the efficiency of data processing.

\section {\label{sec:pot} Potential for femtosecond x-ray nanocrystallography using SASE pulses with  $2
\%$ bandwidth}

In nanocrystallography \cite{CHA3}, crystals of a few hundred
nanometers or smaller are delivered to the interaction region in a
liquid jet. When a crystal in the jet is hit at random and unknown
orientation, such orientation can readily be determined from the
diffraction pattern itself, by indexing the Bragg peaks \cite{KIRI}.
Several auto-indexing programs have been developed for
crystallography. They search for a repeating lattice in the measured
diffraction pattern, knowing the mapping of that pattern onto the
Ewald sphere \cite{OTWI}-\cite{LESL}. In the practical case of a
crystal with different unit-cell spacings in each dimension, the
basis vectors of the reciprocal lattice can be identified quite
readily from the reciprocal lattice spacings observed in the
diffraction pattern. Each Bragg peak in the pattern can be thereby
indexed by its 3D Miller indexes, and properly accumulated. In this
way, diffraction data are build up, averaging over all crystal
shapes and sizes, and yielding complete 3D information as if one
were collecting data from a single undamaged rotating crystal.

\begin{figure}[tb]
\begin{center}
\includegraphics[angle=-90,width=0.75\textwidth]{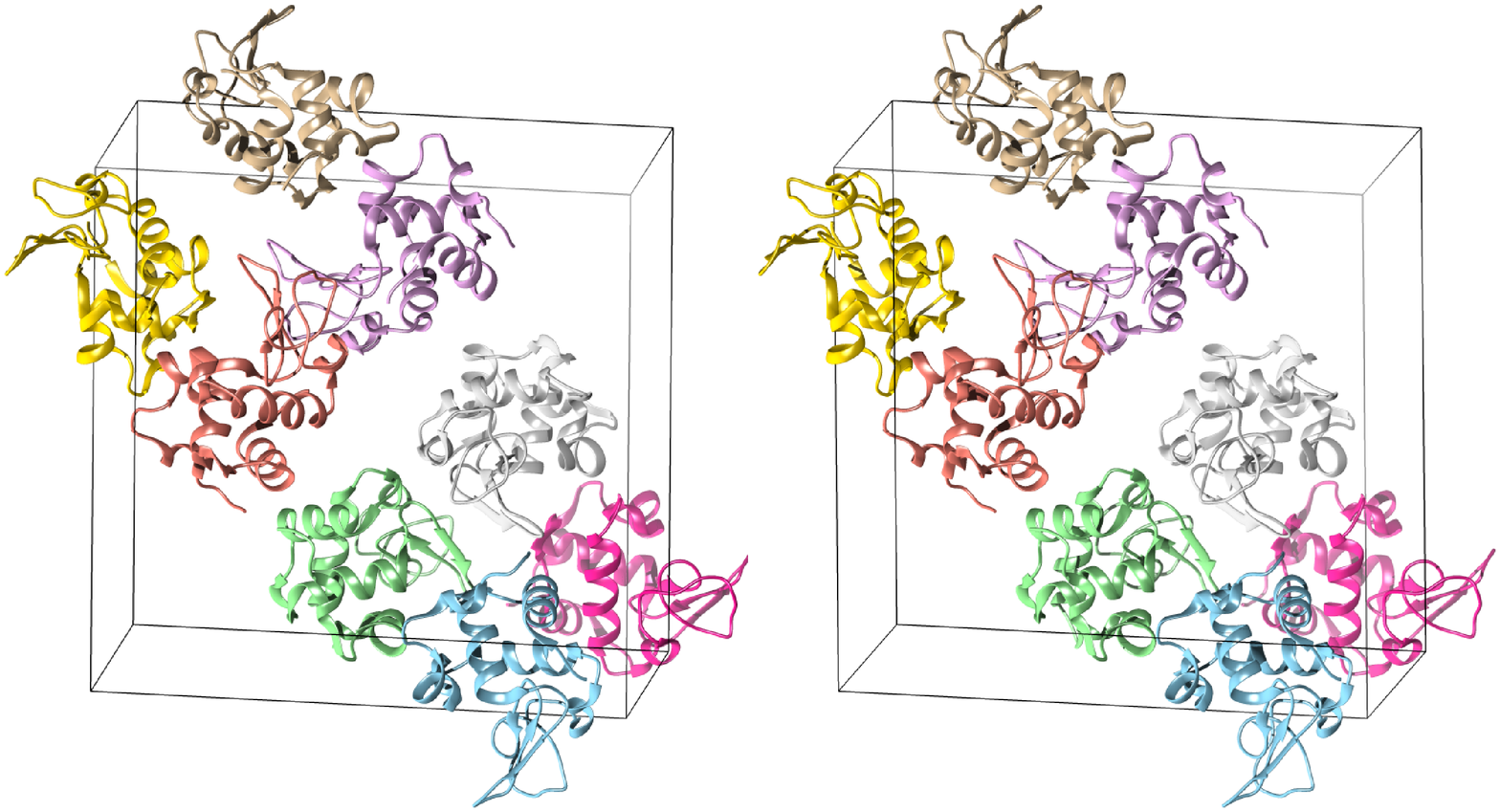}
\end{center}
\caption{Stereo picture of the Lysozyme unit cell, containing $8$
molecules. The cell parameters are: $a=79.18 \AA$, $b=79.18 \AA$,
$c=38.1 \AA$, data from \cite{PDBD}.} \label{lyso}
\end{figure}

Each x-ray pulse produces a diffraction pattern from a single
crystal. The resulting data set consists of thousands of diffraction
patterns from randomly oriented crystals, recorded under "snap-shot
conditions". Each pattern is not angle-integrated across the Bragg
reflections, and is affected by beam divergence, energy spread,
broadening by the small size and possible lattice imperfections of
the crystals. We assume that the pulse duration is sufficiently
short, so that no radiation damage effects occur. Our aim here is to
demonstrate that a tenfold increase in the SASE bandwidth results in
an increase in the number of indexed images at a fixed number of
hits. In this paper we test this concept with simulations of
lysozyme nanocrystals, see Fig. \ref{lyso}. The orientation of each
nanocrystal is determined from the diffraction pattern by using
automated indexing software such as CrystFEL \cite{WHIT}. The
spectral width of the x-ray beam may be simulated by summing
diffracted intensities over a spread of photon energies. For
simulations in this work we neglect the photon beam divergence,
because this effect causes a much smaller effect than the tenfold
increase in bandwidth.

\begin{figure}[tb]
\begin{center}
\includegraphics[clip,width=0.75\textwidth]{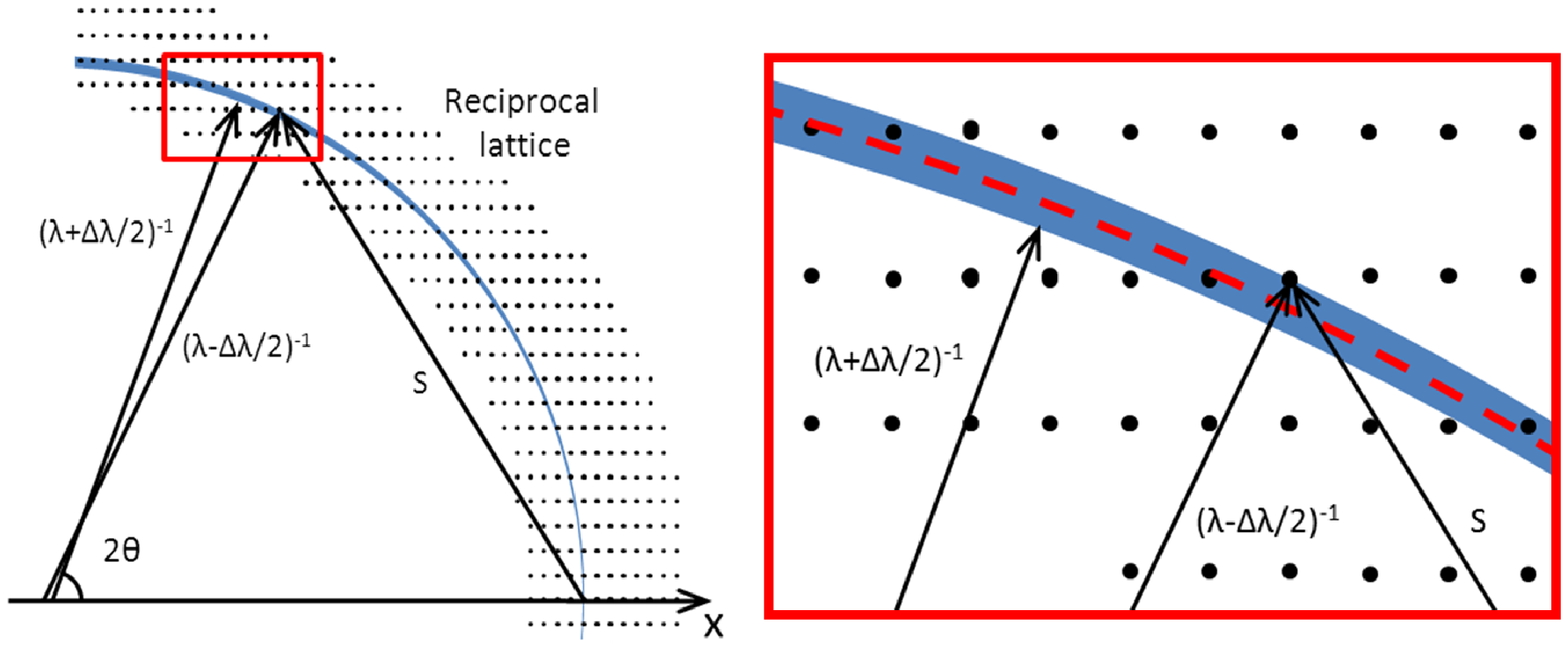}
\end{center}
\caption{Cross section of the reciprocal lattice construction for
imaging of single nanocrystals. The shaded portion of the reciprocal
lattice laying between the two spheres of reflection indicated in
the plot satisfies the diffraction condition for different values of
the wavelength $\lambda$. The thickness of the reflection sphere
results from the finite wavelength bandwidth.} \label{eval}
\end{figure}
Results from our simulated experiment can be represented in terms of
the reciprocal lattice concept, as shown in Fig. \ref{eval}. There
we assume that the spectrum of the incident radiation pulse has a
stepped profile between $\lambda _\mathrm{max}$ and
$\lambda_\mathrm{min}$. Let us consider two Ewald spheres with radii
$1/\lambda_\mathrm{min}$ and $1/\lambda_\mathrm{max}$, tangent to
the origin of the reciprocal lattice, as shown in Fig. \ref{eval}.
The shaded region between them is accessible to the experiment, as
the diffraction condition is satisfied for all reciprocal lattice
points within it. We model the diffraction patterns obtained in this
case as an average over the number $N_p$ of diffraction patterns,
obtained at wavelengths within the bandwidth.

\begin{figure}[htb]
\begin{center}
\includegraphics[clip,width=0.85\textwidth]{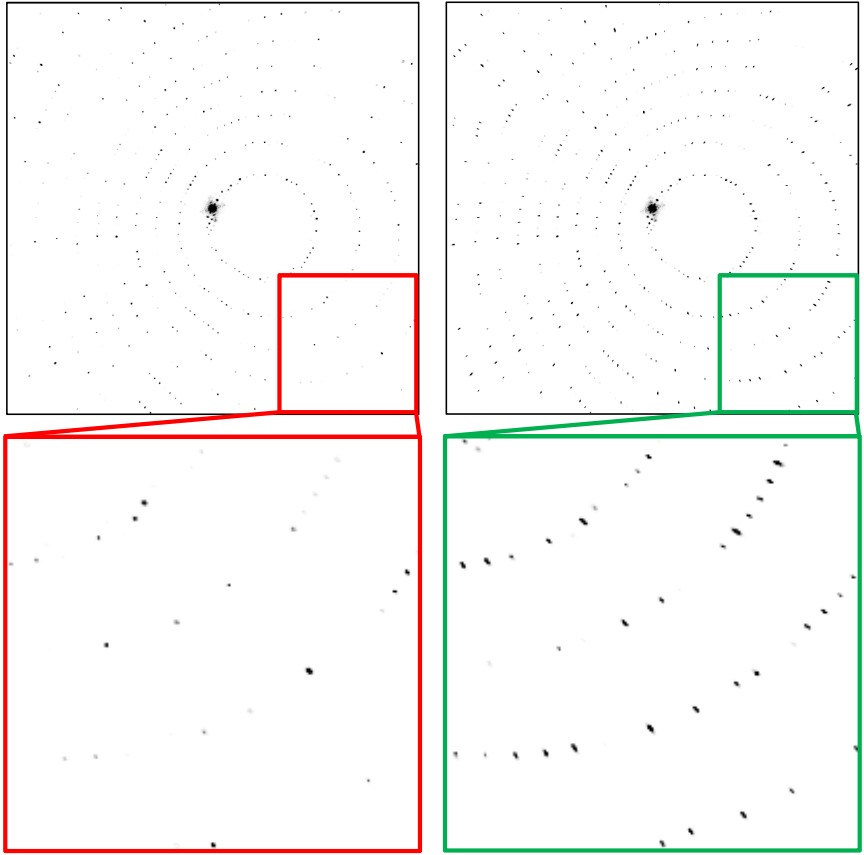}
\end{center}
\caption{Left plot: Diffraction pattern simulated for a single
lysozyme crystal ($20\times 20\times 40$ cells) exposed to an x-ray
pulse with a wavelength of 2.0 $\AA$. The detector considered here
is 176 mm$\times$ 176 mm in size, and is located at 80 mm distance
from the sample. The simulation was performed for 2 mJ photon pulse
energy and focus size of 200 nm. Right: Diffraction pattern
simulated for the same crystal and detector exposed to an X-ray
pulse with finite spectral width, simulated by summing diffracted
intensities  over a spread of wavelengths between
$\lambda_\mathrm{min}= 2.00~ \AA$ and $\lambda_\mathrm{max} = 2.03
~\AA$. The average diffraction patterns, obtained at wavelengths
within the bandwidth is shown for $N_p = 16$. } \label{patt}
\end{figure}
A comparison between diffraction patterns obtained with a
monochromatic and a polychromatic x-ray pulse is shown in Fig.
\ref{patt}. The radiation pulses are short enough (about 3 fs) to
overcome crystal destruction. Therefore, the simulation was
performed neglecting radiation damage. Patterns were calculated for
crystals at random orientations. Fig. \ref{patt} shows a typical
simulation, clearly showing the significant increase in number of
Bragg peaks per individual image in the case of a polychromatic
x-ray beam. Our results suggest that if the bandwidth of the
incident x-ray beam is tenfold increased, the number of indexed
reflections will be much higher.

\section{\label{sec:FELstudy} FEL studies}

We performed a feasibility study with the help of the FEL code
Genesis 1.3 \cite{GENE} running on a parallel machine. We will
present results for the SASE3 FEL line of the European XFEL, based
on a statistical analysis consisting of $100$ runs. The overall
undulator and electron beam parameters used in the simulations are
presented in Table \ref{tt1}.

\begin{table}
\caption{Parameters for the mode of operation at the European XFEL
used in this paper.}

\begin{small}\begin{tabular}{ l c c}
\hline & ~ Units &  ~ \\ \hline
Undulator period      & mm                  & 68     \\
Periods per cell      & -                   & 73   \\
Total number of cells & -                   & 21    \\
Intersection length   & m                   & 1.1   \\
Energy                & GeV                 & 17.5 \\
Charge                & nC                  & 0.25\\
\hline
\end{tabular}\end{small}
\label{tt1}
\end{table}

\begin{figure}[tb]
\includegraphics[width=0.5\textwidth]{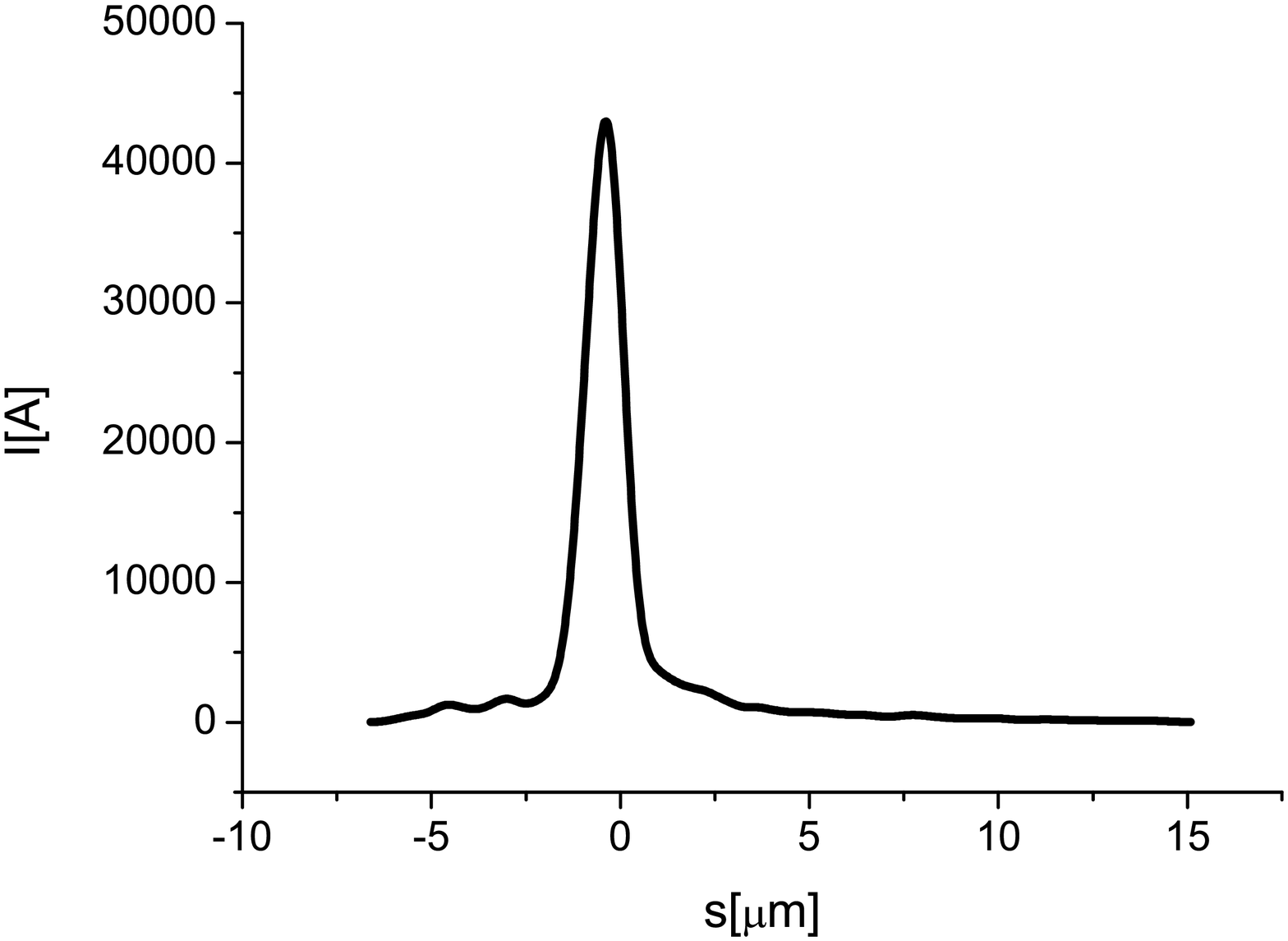}
\includegraphics[width=0.5\textwidth]{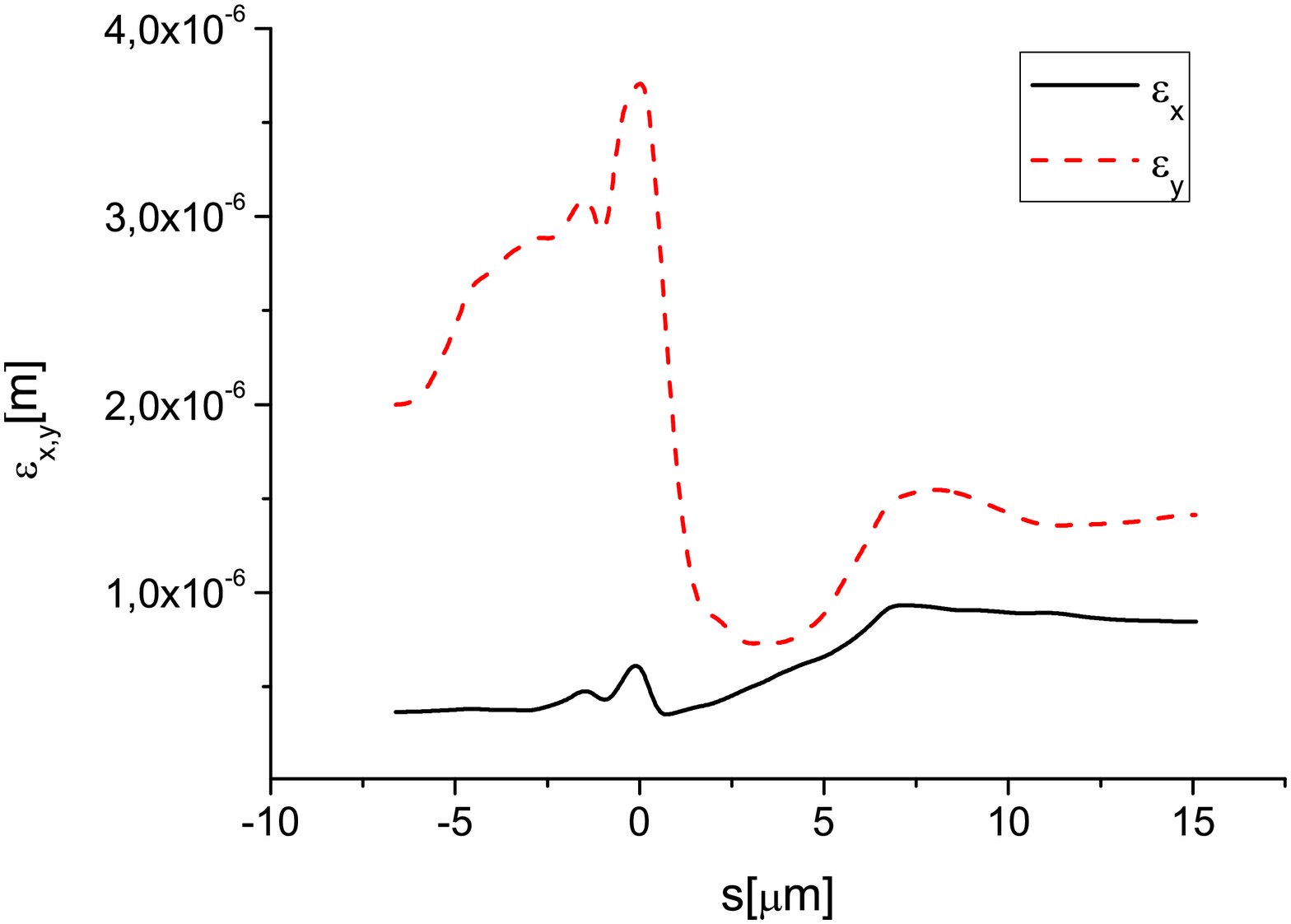}
\includegraphics[width=0.5\textwidth]{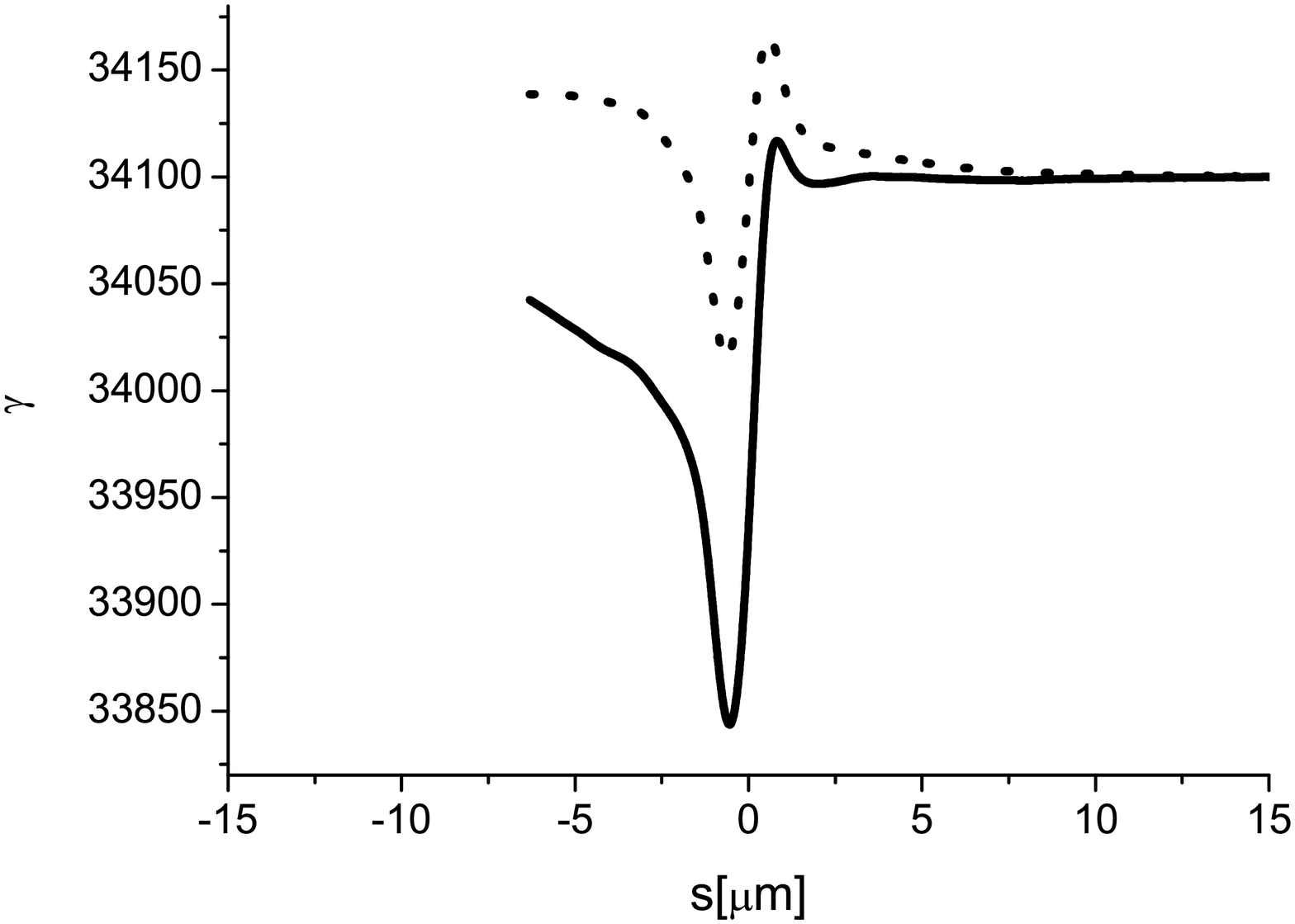}
\includegraphics[width=0.5\textwidth]{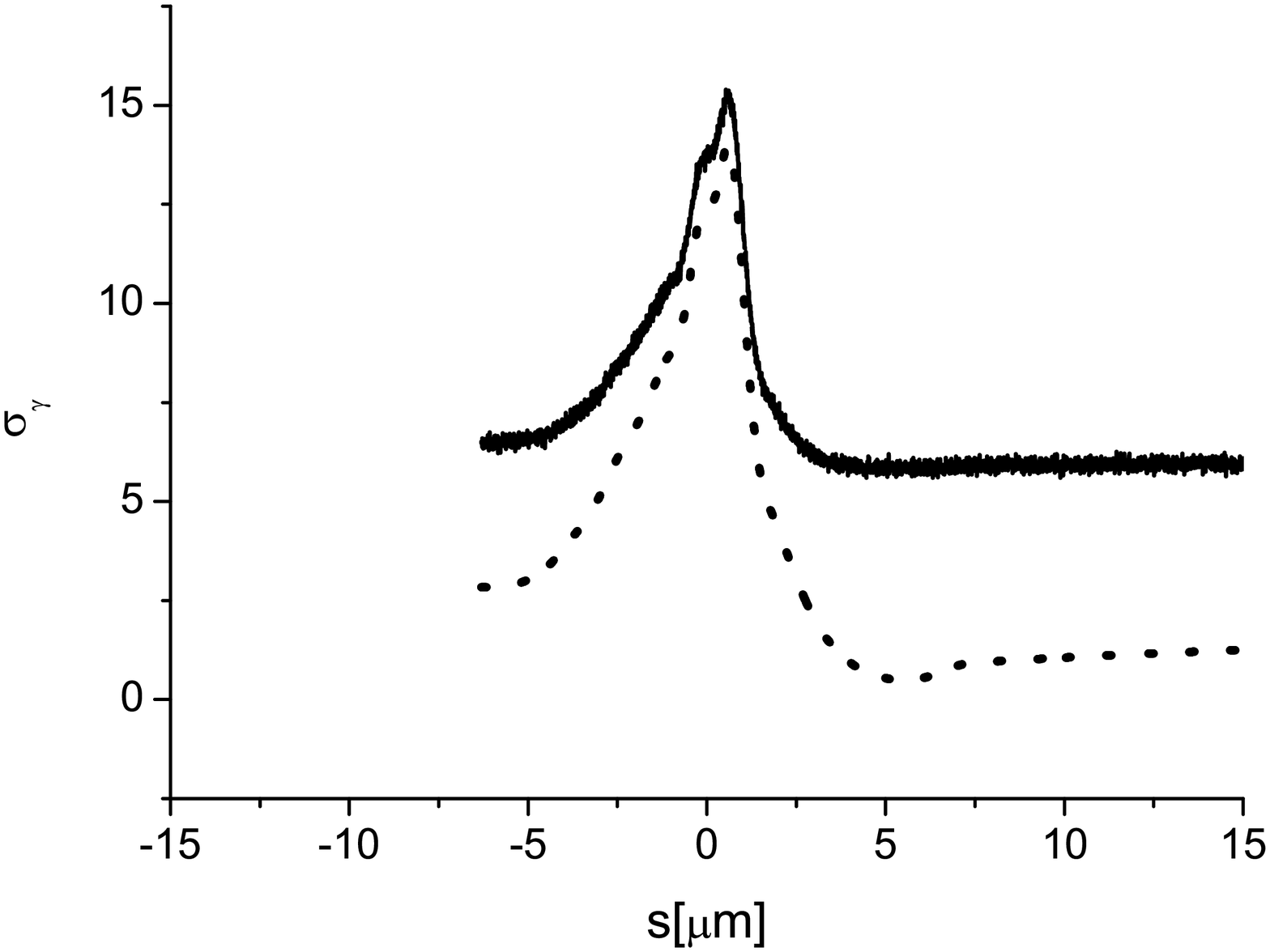}
\begin{center}
\includegraphics[width=0.5\textwidth]{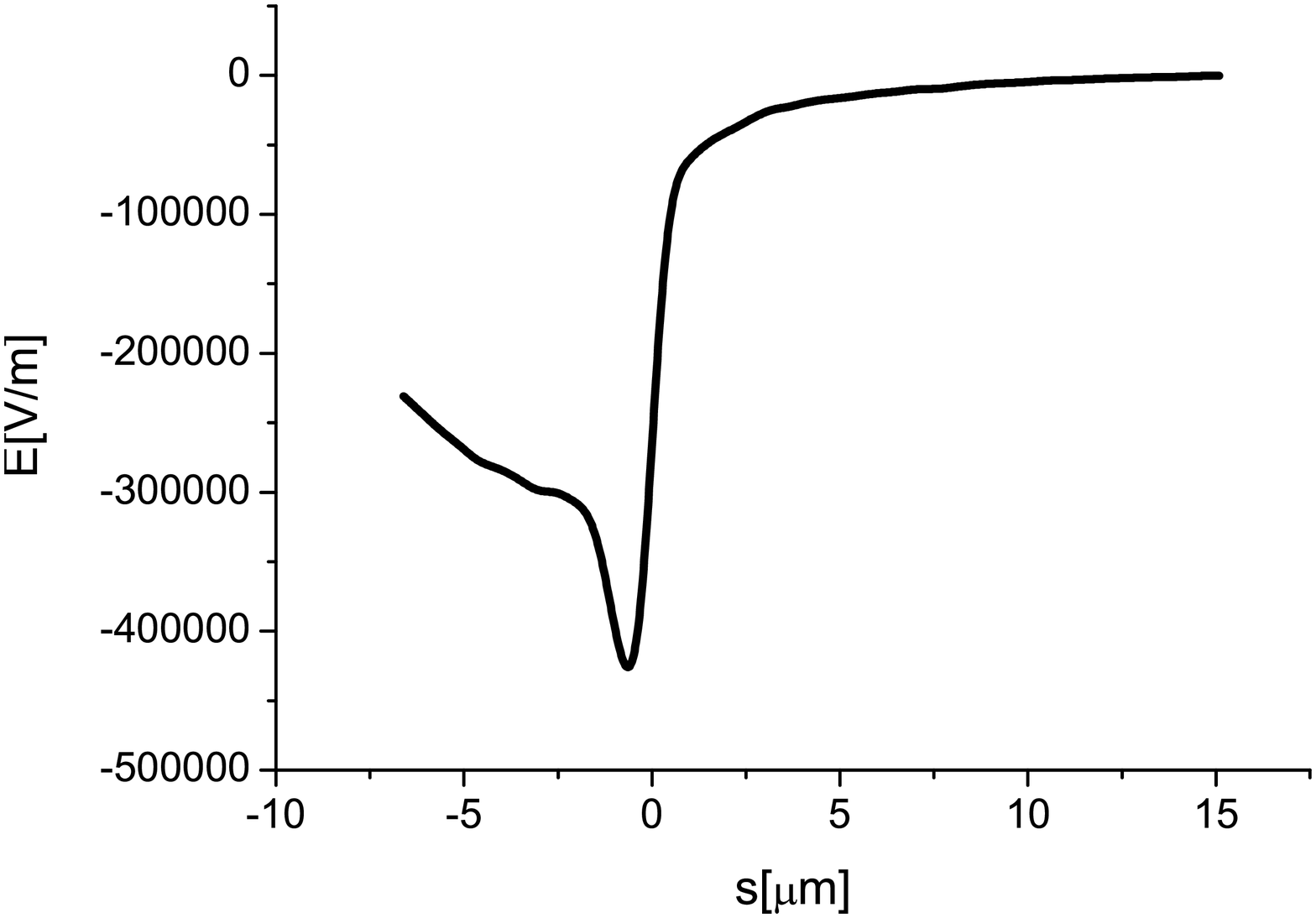}
\end{center}
\caption{Results from electron beam start-to-end simulations. (First
Row, Left) Current profile at the entrance of SASE3. (First Row,
Right) Normalized emittance as a function of the position inside the
electron beam, at the entrance of SASE3. (Second Row, Left) Energy
profile along the beam. Dashed line: at the entrance of SASE1. Solid
line: at the entrance of SASE3. (Second Row, Right) Electron beam
energy spread profile. Dashed line: at the entrance of SASE1. Solid
line: at the entrance of SASE3.  (Bottom row) Resistive wall wake
within the undulator.} \label{s2E}
\end{figure}
\begin{figure}[tb]
\begin{center}
\includegraphics[width=0.5\textwidth]{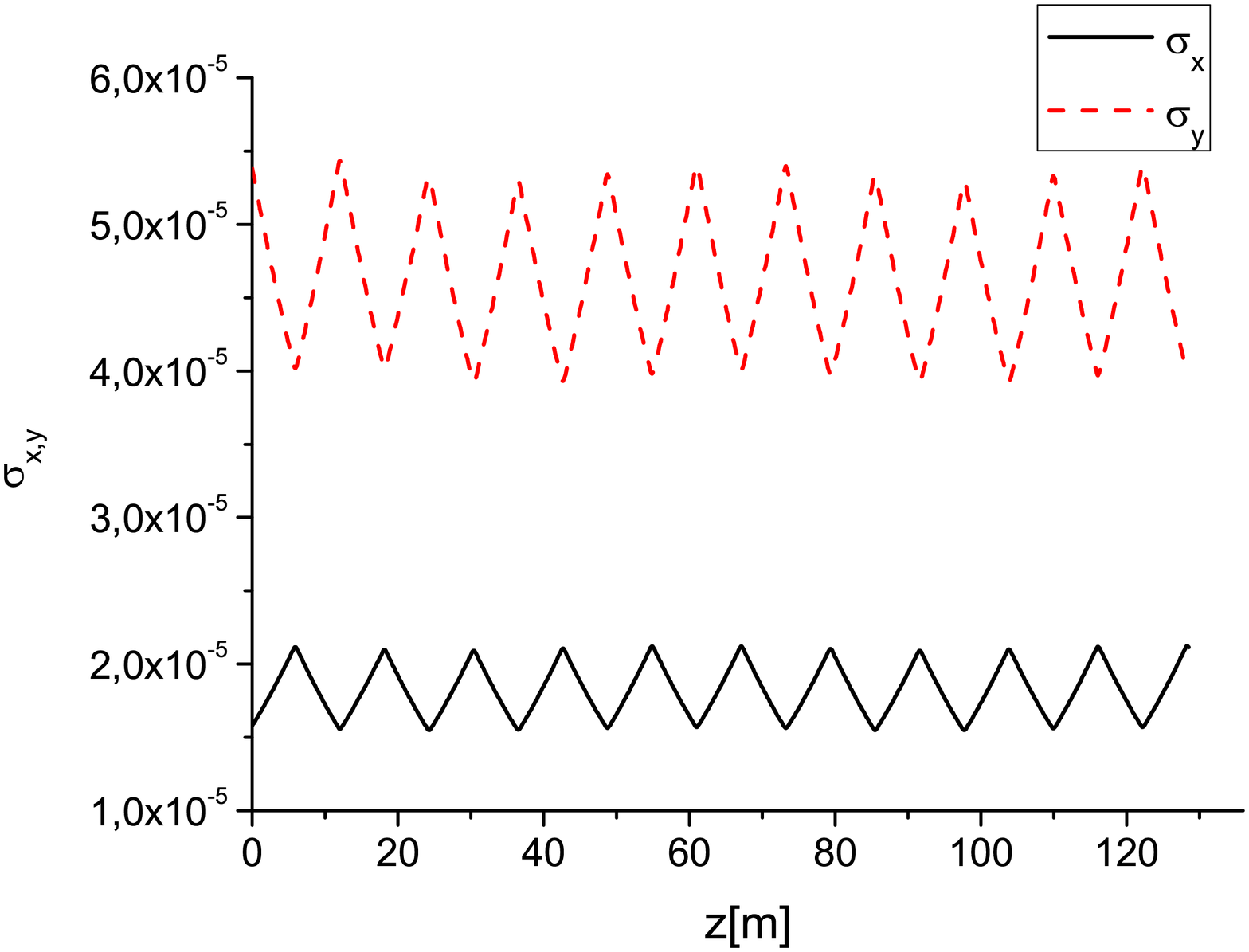}
\end{center}
\caption{Evolution of the horizontal and vertical dimensions of the
electron bunch as a function of the distance inside the SASE3
undulator. The plots refer to the longitudinal position inside the
bunch corresponding to the maximum current value.} \label{sigma}
\end{figure}
The expected beam parameters at the entrance of the SASE3 undulator,
and the resistive wake inside the undulator are shown in Fig.
\ref{s2E}, see \cite{S2ER}. Our calculations account for both wakes
and quantum fluctuations in the SASE1 undulator. The nonlinear
increase of the energy spread is a consequence of the quadratic
superposition of the initial energy spread $\sigma_0(s)$, $s$ being
the coordinate inside the bunch, and the contribution due to quantum
diffusion $\delta$, yielding $\sigma^2(s) = \sigma_0^2(s) +
\delta^2$ . Note the difference in energy chirp and energy spread at
the entrance of SASE1 and at the entrance of SASE3. The additional
energy chirp induced during the passage through SASE1, due to the
presence of resistive wakes, is very helpful in our case, because it
increases the energy chirp, and finally leads to a larger bandwidth.

Due to collective effects in the bunch compression system,
emittances in the horizontal and vertical directions are
significantly different. As a result, the electron beam looks highly
asymmetric in the transverse plane: in the horizontal direction
$\sigma_x \sim 20 \mu$m, while in the vertical direction $\sigma_y
\sim 50 \mu$m. The evolution of the transverse electron bunch
dimensions are plotted in Fig. \ref{sigma}.

\begin{figure}[tb]
\includegraphics[width=0.5\textwidth]{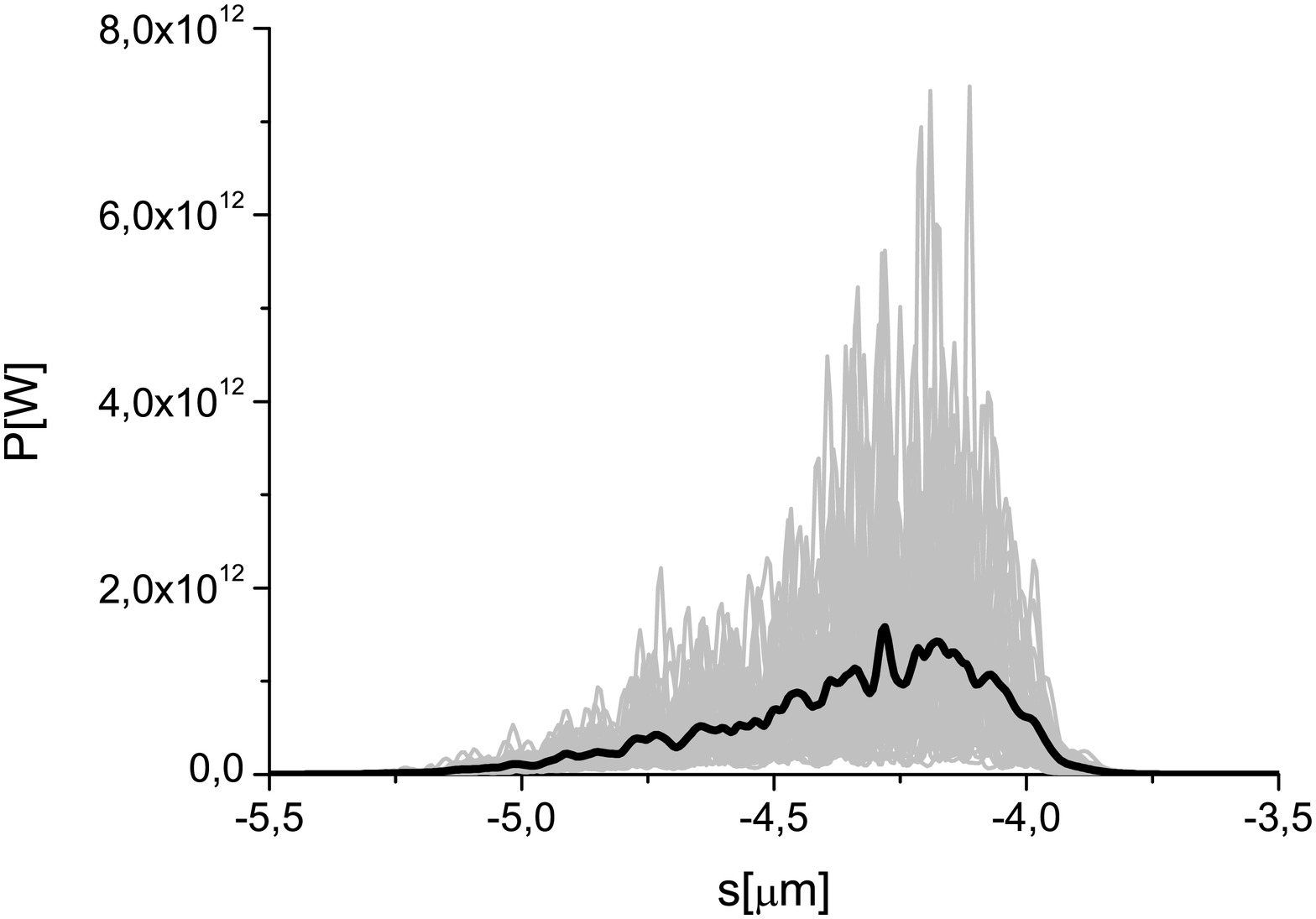}
\includegraphics[width=0.5\textwidth]{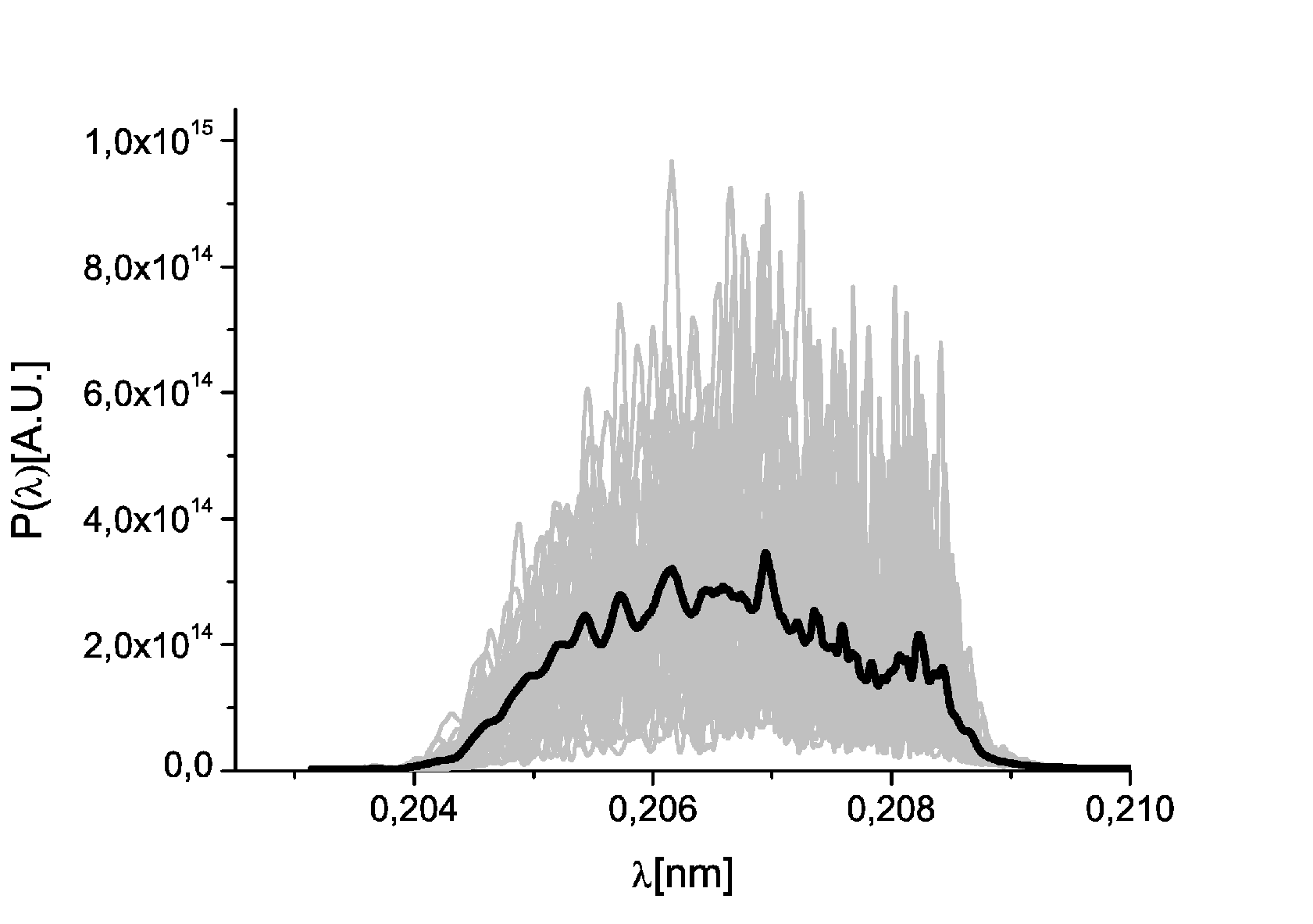}
\caption{Power distribution  and spectrum of the X-ray radiation
pulse after the SASE3 undulator. Grey lines refer to single shot
realizations, the black line refers to the average over a hundred
realizations.} \label{PSP}
\end{figure}
\begin{figure}[tb]
\includegraphics[width=0.5\textwidth]{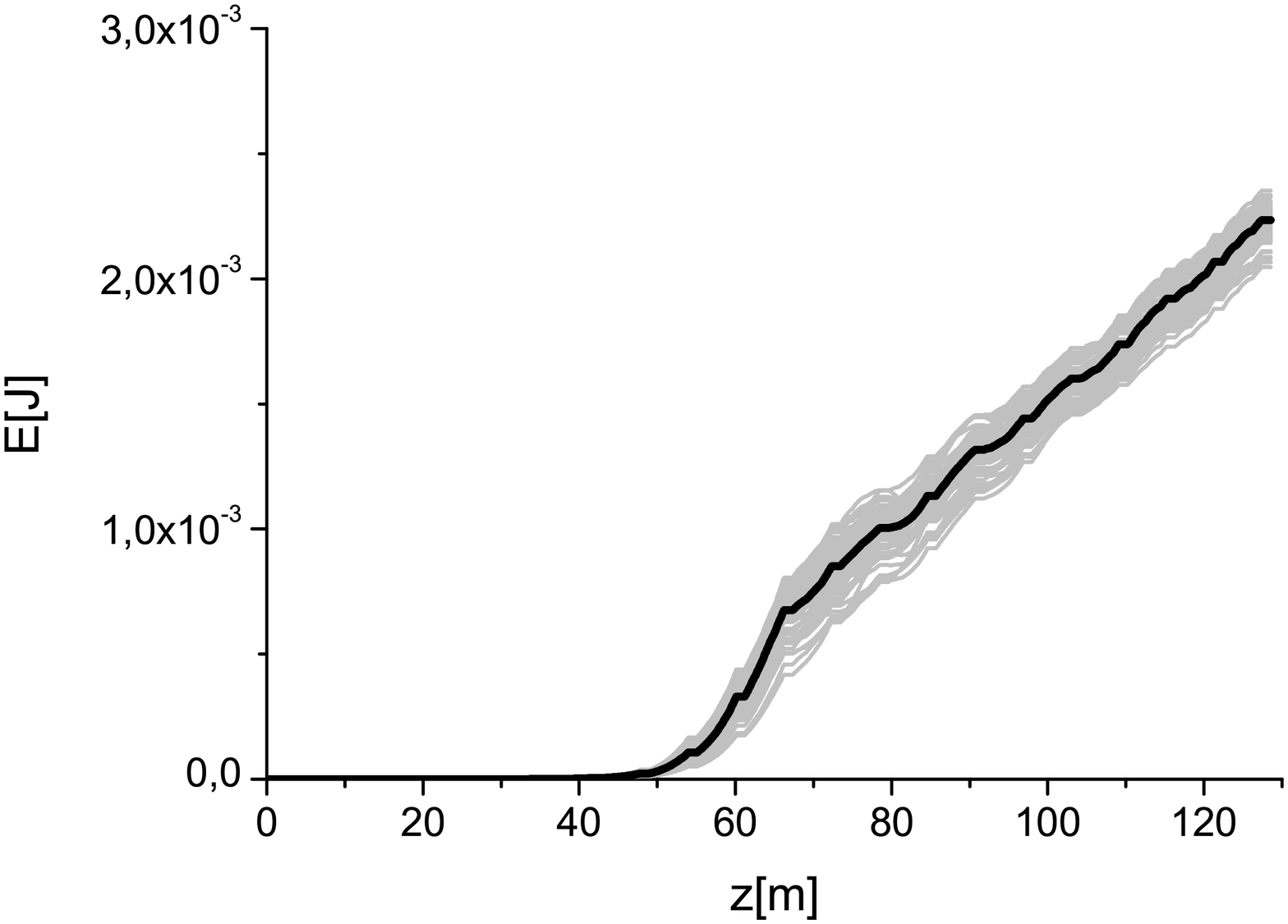}
\includegraphics[width=0.5\textwidth]{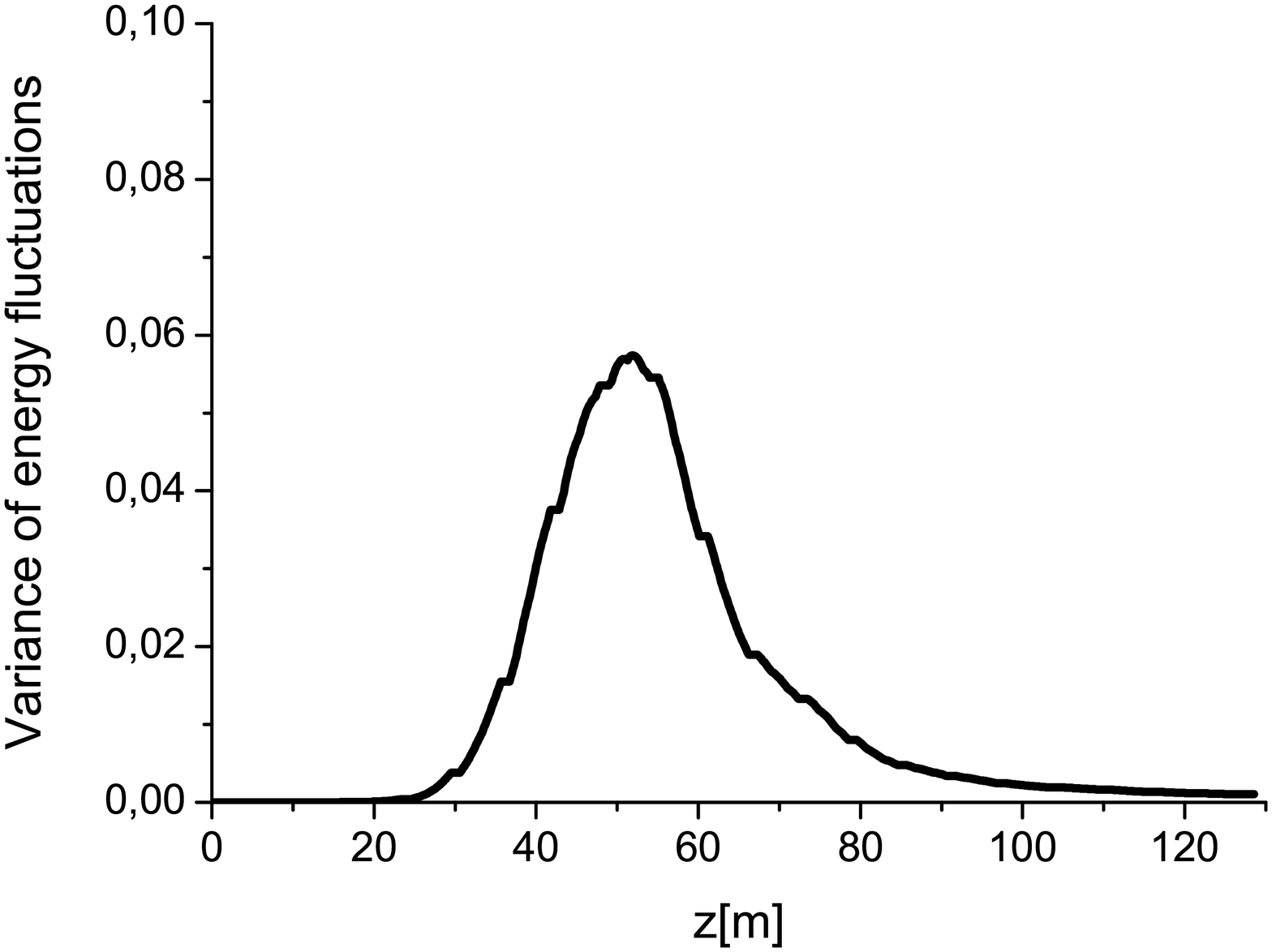}
\caption{Evolution of the energy per pulse and of the energy
fluctuations as a function of the undulator length. Grey lines refer
to single shot realizations, the black line refers to the average
over a hundred realizations.} \label{OUT2}
\end{figure}
\begin{figure}[tb]
\includegraphics[width=0.5\textwidth]{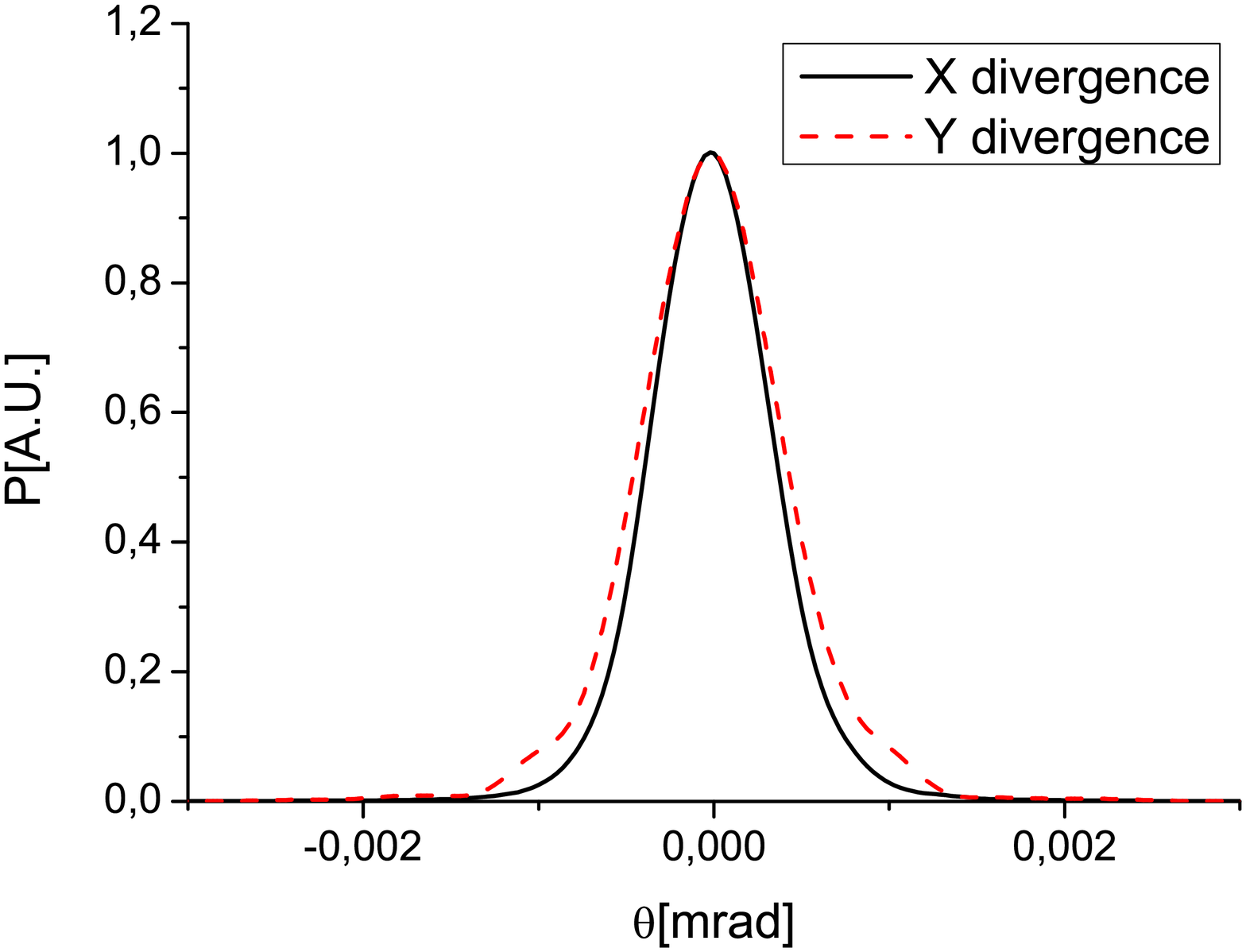}
\includegraphics[width=0.5\textwidth]{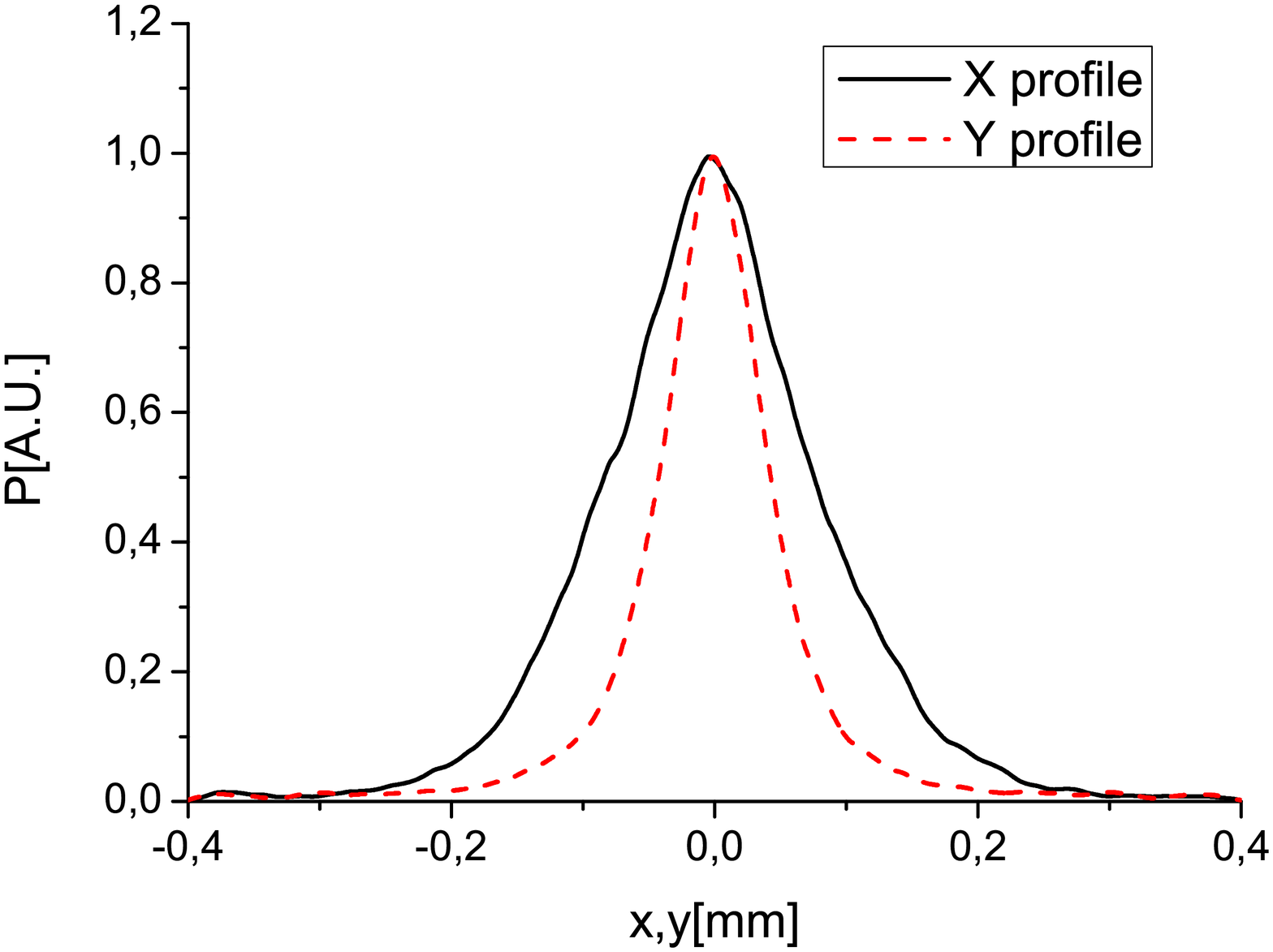}
\caption{(Left plot) Distribution of the radiation pulse energy per
unit surface and (right plot) angular distribution of the radiation
pulse energy at the exit of the SASE3 undulator.} \label{spotT}
\end{figure}
The output SASE power distribution and spectrum after the $21$
undulator segments of SASE3 are shown in Fig. \ref{PSP}. The
evolution of the energy per pulse and of the energy fluctuations as
a function of the undulator length are shown in Fig. \ref{OUT2}.
Finally, the distribution of the radiation pulse energy per unit
surface and the angular distribution of the radiation pulse energy
at the exit of SASE3 undulator are shown in Fig. \ref{spotT}.

Note that, since the electron beam is transversely asymmetric, the
output radiation beam is asymmetric too. Intuition would incorrectly
suggest that a larger size of the electron bunch should be
responsible for a larger size of the radiation beam, but a quick
comparison of Fig. \ref{sigma} and Fig. \ref{spotT} show that this
is not the case in reality. In fact we see that, in the vertical
direction, the size of the photon beam is about $50 ~\mu$m,
practically the same as the electron beam. This means that, within a
gain length, the expansion of the radiation beam due to diffraction
in the vertical direction is much less than the vertical size of the
electron bunch. On the contrary, in the other direction radiation
expands outside of the electron bunch due to diffraction effects,
and the photon beam size grows to about $100 ~\mu$m, which is $5$
times larger than the electron bunch size.

Finally, it should be appreciated that the characteristics of the
radiation beam produced as described in this article are very
suitable for nanocrystallography. The photon energy is about $6$
keV, and radiation is delivered in $3$ fs pulses with an energy of
about $2.5$ mJ . Our study shows that for the particular mode of
operation investigated here, we have straightforwardly $1$ TW power
in the SASE regime already, which constitutes an increase of about
two orders of magnitude compared with the nominal mode of operation.

\section{\label{sec:cons} Conclusions}

Output characteristics of the European XFEL have been previously
studied assuming an operation point at 5 kA peak current. In this
case, the baseline SASE undulator sources will saturate at about 50
GW (see e.g. \cite{TSCH}). This power limit is very far from the
multi-TW level required for bio-imaging applications. In this paper
we explore the possibility to go well beyond the nominal 5 kA peak
current. In order to illustrate the potential of this approach we
consider a 0.25 nC bunch compressed up to 45 kA peak current. Based
on start-to-end simulations it is shown here that 2 TW power could
be generated in the SASE regime for a photon energy range between 2
keV and 6 keV, which is optimal for femtosecond X-ray
nanocrystallography and single biomolecular
imaging\footnote{Numerical calculations show that, using undulator
tapering in the SASE mode of operation, it is possible to reach 2 TW
peak power at 6 keV. However, such undulator tapering additionally
reduces the bandwidth from $2 \%$ in an uniform undulator to about
$1 \%$. Since our goal here is to maximize the bandwidth for
applications in nanocrystallography, in this article we consider the
uniform undulator option, allowing for generation of radiation
pulses of 1 TW power and $2 \%$ bandwidth.}. This example
illustrates the potential for improving the performance of the
European XFEL without additional hardware\footnote{Our simulations
are based on the 21 cells foreseen for the SASE3 undulator beamline
\cite{TSCH}, Table \ref{tt1}.}. This solution to generate TW power
mode of operation is not without complexities. The price for using a
very high peak-current is operation with a large energy chirp within
the electron bunch, yielding a large SASE radiation bandwidth.
However, it is shown here that there are applications like
nanocrystallography, where x-ray radiation pulses with a few percent
bandwidth present many advantages, compared to those produced in the
nominal SASE mode of operation, which have a relative bandwidth of a
fraction of a percent.

\section{Acknowledgements}

We are grateful to Massimo Altarelli, Reinhard Brinkmann, Henry
Chapman, Janos Hajdu, Viktor Lamzin, Serguei Molodtsov and Edgar
Weckert for their support and their interest during the compilation
of this work. We thank Edgar Weckert for providing the code MOLTRANS
to one of us (O.Y.).

\end{document}